\documentclass[aps,prl,twocolumn,superscriptaddress,showpacs,floatfix]{revtex4-2}

\usepackage{amsmath,amssymb,amsfonts,float,graphics,epsfig,epstopdf,color,verbatim,tabularx,bm,multirow,appendix}
\usepackage[utf8]{inputenc}
\usepackage[T1]{fontenc}
\usepackage{xcolor}

\usepackage{dsfont}
\usepackage{textcomp}
\usepackage{yfonts}
\usepackage{bm}
\usepackage{subfigure}
\usepackage{mathrsfs}
\usepackage{graphicx}
\usepackage{verbatim}
\usepackage{hyperref}
\usepackage{braket}
\usepackage[normalem]{ulem}
\usepackage{tikz}
	\usetikzlibrary{calc}
	\usetikzlibrary{shapes.multipart}
\usepackage{orcidlink}

\makeatletter
\def\l@subsubsection#1#2{}
\makeatother

\DeclareMathOperator{\Tr}{Tr}

\begin{document}

\title{Criticality on R\'enyi Defects at (2+1)$D$ O(3) Quantum Critical Points}
\author{Yanzhang Zhu\,\orcidlink{0009-0001-8264-6993}}
\affiliation{Department of Physics, School of Science and Research Center for Industries of the Future, Westlake University, Hangzhou 310030, China}
\affiliation{Institute of Natural Sciences, Westlake Institute for Advanced Study, Hangzhou 310024, China}
\affiliation{State Key Laboratory of Surface Physics and Department of Physics, Fudan University, Shanghai 200433, China}
\author{Zhe Wang\,\orcidlink{0009-0007-6501-5421}}
\affiliation{Department of Physics, School of Science and Research Center for
Industries of the Future, Westlake University, Hangzhou 310030, China}
\affiliation{Institute of Natural Sciences, Westlake Institute for Advanced Study, Hangzhou 310024, China}
\author{Meng Cheng\,\orcidlink{0000-0001-7600-1809}}
\email{m.cheng@yale.edu}
\affiliation{Department of Physics, Yale University, New Haven, Connecticut 06511, USA}
\author{Zheng Yan\,\orcidlink{0000-0002-3349-5965}}\email{zhengyan@westlake.edu.cn}
\affiliation{Department of Physics, School of Science and Research Center for
Industries of the Future, Westlake University, Hangzhou 310030, China}
\affiliation{Institute of Natural Sciences, Westlake Institute for Advanced Study, Hangzhou 310024, China}
\date{\today}

\begin{abstract}
At a quantum critical point, the universal scaling behavior of R\'enyi entanglement entropy is controlled by the universality class of the codimension-two R\'enyi (or conical) defects in the infrared theory. In this Letter we perform a systematic study of critical correlations along R\'enyi defect lines in (2+1)d quantum spin models realizing quantum phase transitions described by the O(3) Wilson-Fisher universality class, using large-scale quantum Monte Carlo simulations. We present numerical evidence that, for a fixed R\'enyi index $n$, there exist multiple R\'enyi defect universality classes, with distinct critical exponents for the O(3) order parameter on the defect. These universality classes are realized by choosing microscopically different entanglement cuts in lattice models, which we classify as ordinary, special, and extraordinary according to their relation to surface criticality. For the extraordinary entanglement cut, we further find  evidence for a phase transition on the defect as a function of the R\'enyi index. Our results highlight the key role of defect universality classes in determining the universal scaling of R\'enyi entanglement entropy, and provide a framework for understanding the previously observed dependence of R\'enyi entanglement entropy scaling on microscopic lattice details.
\end{abstract}

\maketitle

\textit{\color{blue}Introduction ---}
Characterizing the structure of quantum entanglement is a key task to understand quantum many-body states. Bipartite entanglement can be measured through the entanglement entropy (EE) of one subsystem, and a commonly used measure is the R\'enyi EE~\cite{calabrese2004entanglement,amico2008entanglement,eisert2010colloquium,calabrese2009entanglement,nishioka2018entanglement,holzhey1994geometric,korepin2004universality,laflorencie2016quantum,kallin2011anomalies,metlitski2009entanglement,metlitski2011entanglement,li2008entanglement,kitaev2006topological}. In quantum many-body systems, the size and shape dependence of R\'enyi EEs reveal important information about the underlying state. A well-known example is the logarithmic dependence of R\'enyi EE on subsystem size in (1+1)D conformal field theories (CFTs), with the coefficient fixed by the central charge~\cite{calabrese2004entanglement, calabrese2009entanglement}. Thus entanglement measurement has become a standard tool to extract the central charge at (1+1)D quantum critical points (QCPs). Similarly, universal subleading corrections have been uncovered in Goldstone phases and QCPs in (2+1)D or higher~\cite{casini2007universal, fradkin2006entanglement,metlitski2009entanglement,metlitski2011entanglement,deng2024diagnosing,lin2007entanglement,kallin2011anomalies,kallin2014corner,helmes2014entanglement,park2015logarithmic, WhitsittEE,  BuenoPRL2015, HelmesPRB2016, zhao2022scaling, song2024extracting}. Recently, advances in numerical algorithms have enabled high-precision measurements of R\'enyi EE in large-scale quantum Monte Carlo (QMC) simulations in higher dimensions~\cite{hastings2010measuring,luitz2014improving,d2020entanglement,grover2013entanglement,humeniuk2012quantum,kallin2013entanglement,kallin2011anomalies,helmes2014entanglement, zhao2022measuring, da2024extracting, wang2025universal, wang2025bipartite}. This, in turn, allows access to universal subleading terms in the scaling behavior.

However, it has also been observed in recent studies that the presumably universal quantities of entanglement are actually sensitive to the microscopic details of the entanglement cut~\cite{zhao2022scaling, wang2026universal, REE_GN}. For example, the coefficient of the corner-induced logarithmic term in R\'enyi-2 EE at QCPs, presumably a universal quantity, is found to depend on how the cuts are defined on the lattice in several different critical lattice models~\cite{d2024entanglement,deng2024diagnosing,zhu2026bipartite,zhao2022scaling,REE_GN, liu2024demonstrating}. This puzzling phenomenon points to an incomplete understanding of how microscopic realizations of entanglement cuts are encoded in the infrared (IR) field-theoretical description of R\'enyi EE.

In this Letter, we approach this important problem from the perspective of defect universality classes.  In a field-theoretical framework, the R\'enyi EE of a region $A$ is expressed as the free energy of the system in the presence of a ``R\'enyi defect'' along the boundary of $A$~\cite{bianchi2016renyi}. Such an extended defect, embedded in the bulk CFT, has its own universality class (or phases), characterized by fixed points of defect renormalization group (RG) flows (while the bulk remains critical)~\cite{andrei2020boundary,cuomo2022renormalization,wang2026boundary}. A familiar example is an impurity or a pinning field, which gives rise to a (0+1)D defect in spacetime~\cite{affleck2000conformal,cuomo2022localized,wu2026impurity}. Physically, correlation functions of local observables can exhibit long-distance behavior distinct from those in the bulk.

The defect universality classes determine both local (e.g. long-distance behavior of local observables) and global properties, including the geometry/shape dependence of R\'enyi EE. Importantly, for a given R\'enyi index, there may exist multiple distinct defect phases in the IR~\cite{metlitski2009entanglement}. If this occurs, microscopically different entanglement cuts can flow to distinct defect universality classes, resulting in different shape dependence of R\'enyi EE.

Guided by this observation, in this Letter we systematically investigate the critical behavior of local observables on R\'enyi defects in (2+1)D quantum spin models realizing O(3) QCP, using QMC simulations on the R\'enyi manifolds. This approach allows for a more direct probe of defect phases, and is complementary to the more ``global'' measurements of defect entropy commonly studied in recent works.
We present numerical evidence that the so-called ordinary, special, and extraordinary entanglement cuts  introduced below realize distinct R\'enyi-defect universality classes, and that the
extraordinary cut undergoes a phase transition as the R\'enyi index is varied.

\textit{\color{blue}Model and methods---} In this Letter we study quantum spin models realizing O(3) QCP, focusing on the spin-$1/2$ columnar dimerized Heisenberg model on the square lattice. The Hamiltonian is
\begin{equation}
H = J \sum_{\langle ij\rangle} \mathbf{S}_i \cdot \mathbf{S}_j + J' \sum_{\langle ij\rangle'} \mathbf{S}_i \cdot \mathbf{S}_j ,
\end{equation}
where $\langle ij\rangle$ and $\langle ij\rangle'$ denote weak and strong bonds, respectively. We tune the coupling ratio to the O(3) bulk QCP at $J'/J = 1.9096$. This model has been widely used to investigate surface critical behavior under different physical boundary cuts~\cite{ding2018engineering}. In the present work, we use different entanglement bipartitions of the same lattice to realize distinct R\'enyi defects. For comparison, we have also studied the bilayer Heisenberg model and the staggered dimerized Heisenberg model, finding consistent results for the defect criticality; these results are presented in the Supplemental Material (SM)~\cite{supplemental}. For all cases, in the bulk the transition between the N\'eel phase and dimerized phase is described by the O(3) Wilson-Fisher CFT.

\begin{figure}[ht]
    \centering
    \includegraphics[width=1.0\linewidth]{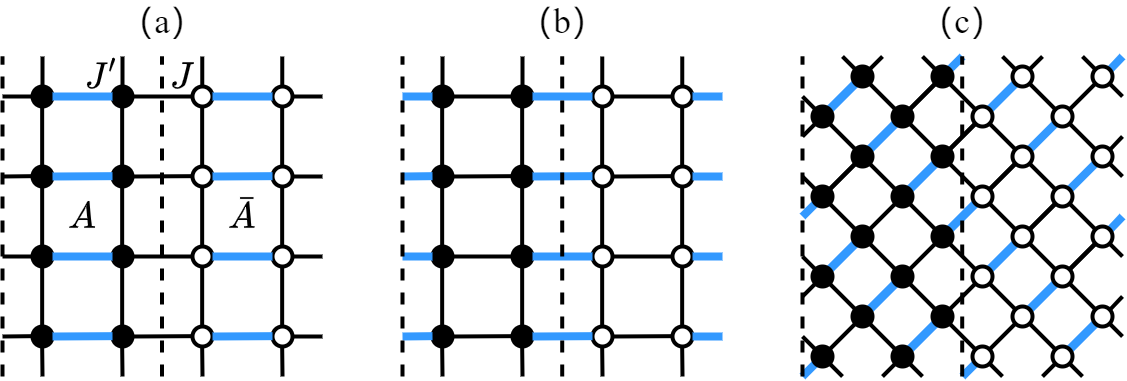}
    \caption{Schematic illustration of the columnar dimerized Heisenberg model and entanglement bipartitions. Thick blue bonds represent the strong couplings $J'$, and thin black bonds denote the weak couplings $J$. The dashed line indicates the entanglement bipartition, which separates the system into subsystem $A$ (in black) and its environment $\bar{A}$ (in white). Panels (a) and (b) correspond to ordinary and special cuts, while panel (c) corresponds to the extraordinary cut realized by a tilted cut in the same model. These three cuts are used throughout the Letter to realize the three classes of R\'enyi defects compared below.}
    \label{fig:model_dim}
\end{figure}

We consider the entanglement bipartition of a system with periodic boundary conditions. More specifically, the lattice is divided into two subsystems along a straight cut, and one subsystem is traced out. This construction generates a R\'enyi-$n$ defect, which forms a one-dimensional structure along the entanglement bipartition edge. The R\'enyi index $n$ can be interpreted as an effective inverse temperature for the entanglement Hamiltonian. The R\'enyi defect can be represented by a $n$-sheeted replica manifold and simulated directly within QMC~\cite{yan2021entanglement,li2024relevant,song2023different,wu2023classical,liu2026worldline}. Since the bulk critical point has dynamical exponent $z=1$, we set the inverse temperature proportional to the system size, $\beta=2L$, to access the ground-state regime. For sites in subsystem $A$, the imaginary-time boundary at $\tau=\beta$ of replica $k$ is connected to $\tau=0$ of replica $(k+1) \bmod n$, while sites in $\bar A$ obey the usual periodic imaginary-time boundary condition within each replica. The largest system size used in the main simulations is $L=80$, and statistical errors are estimated from independent Monte Carlo bins.

A schematic illustration of the dimerized Heisenberg model with different entanglement bipartitions is shown in Fig.~\ref{fig:model_dim}. 
We borrow the names ordinary [Fig.~\ref{fig:model_dim}(a)], special [Fig.~\ref{fig:model_dim}(b)], and extraordinary [Fig.~\ref{fig:model_dim}(c)] from the
corresponding boundary criticality problems: if the system is physically cut
along these lines, the resulting boundaries realize the surface
universality classes with the same names
~\cite{cardy1996scaling, binder1974surface, deng2005surface,  zhang2017unconventional,ding2018engineering, weber2018nonordinary,ding2023special,wang2024surface,wang2026detecting}. More specifically, both ordinary and special boundaries remain disordered, and the extraordinary boundary develops long-range ferro- or ferrimagnetic order~\cite{ding2018engineering,wang2024surface,wang2026detecting}. 

To characterize the properties of the R\'enyi defect, we measure spin correlations and Binder cumulants defined on the $n$-sheeted replica manifold for each type of entanglement cut. For an observable $O$ defined on subsystem $A$, we define the R\'enyi-$n$ expectation value as
\begin{equation}
    \braket{O}_n = \frac{\Tr (O\rho_A^n)}{\Tr \rho_A^n}.
\end{equation}
In field theory, one can think of $\braket{O}_n$ as the expectation value of $O$ in the presence of a R\'enyi-$n$ defect. We are mostly interested in $O$ along the entanglement cut, so $\braket{O}_n$ allows us to access local observables on the defect. In practice, measuring correlations of local observables on R\'enyi defects in QMC simulations is computationally much more efficient than measuring the R\'enyi EE itself.

In this way, we compute the equal-time defect spin-spin correlation $C_s(L)$ between two surface spins $i$ and $j$ with the longest distance $|i-j|=L/2$:
 $C_s(L)=\braket{S^z_i S^z_{i+L/2}}_n$.
Long-range magnetic order is indicated by a finite value of $C_s(L)$ in the thermodynamic limit $L \to \infty$. We also evaluate the defect Binder cumulant 
 $ U_{2}(L)= \frac{5}{6}\left(3-\frac{\langle M_z^{4}\rangle_n}{\langle M_z^{2}\rangle_n^{2}}\right)$,  
where the magnetization $M_z =\sum_{i \in {\rm cut}} \eta_i S_i^z$, $\eta_i = \pm 1$ depending on which sublattice spin $i$ belongs to. Notice that $\eta_i$ are alternating signs for ordinary and special cuts, hence $M_z$ measures \emph{antiferromagnetic} order. On the other hand, for the extraordinary cut $\eta_i$ are all the same along the line (see Fig.~\ref{fig:model_dim}).
In the ordered phase, $U_2(L) \rightarrow 1$ in the thermodynamic limit $L \to \infty$. In the disordered phase, $U_2$ approaches a value between 0 and 1.

For disordered states, the finite-size behavior of the correlation function is well described by a power law $C_s(L) \sim L^{-2\hat{\Delta}_{\bm{\phi}}}$ with subleading corrections. In practice, we include correction-to-scaling terms to account for finite-size corrections and obtain stable estimates of the exponent for accessible system sizes. Thus the fitting formula is  
\begin{equation}
\label{eq:CsLfit}
    C_s(L)=aL^{-2\hat{\Delta}_{\bm{\phi}}}(1+bL^{-1}).
\end{equation}

In the ordered regime, the correlation function approaches a finite value in the thermodynamic limit. In this case, the leading behavior is a finite constant, so we instead use polynomial fits in $1/L$ to extrapolate the asymptotic value, which provide a more reliable description of the finite-size behavior.

\textit{\color{blue}R\'enyi defect CFT---}  We now connect the observables measured in numerical simulations to those in the underlying field theory. The bulk of the system near the transition is well described by the 3D O(3) Wilson-Fisher CFT, and we will denote the fundamental order-parameter field by an O(3) vector $\bm{\phi}(\bm{x})$, with scaling dimension $\Delta_{\bm{\phi}}\approx 0.52$. It is well-understood that in the bulk $\bm{\phi}$ is the continuum limit of the spin operator $\bm{S}_i$ \footnote{$\bm{S}$ also contains the density component of the O(3) current. But it is more irrelevant ($\Delta_J=2$) than the order parameter ($\Delta_{\bm{\phi}}\approx 0.52$). These two operators also appear with distinct sublattice sign patterns.}.

The R\'enyi EE of order $n$ is given by the free energy of the theory with the insertion of the R\'enyi defect (also known as conical singularity) along the boundary of the entangling region. Around the defect, the field satisfies the following boundary condition:
\begin{equation}
    \bm{\phi}(r, \theta, x_\perp)=\bm{\phi}(r, \theta+2\pi n, x_\perp),
\end{equation}
where $r$ and $\theta$ are the local polar coordinates on the plane transverse of the defect, and $x_\perp$ is the coordinate along the defect. Note that $n$ can be any real number, but numerically we only have access to positive integer $n>1$ by the replica QMC~\cite{yan2021entanglement}.

The IR limit of the defect can be described by observables localized on the defect \cite{billo2013line,billo2016defects}. In particular, when $\bm{\phi}$ is brought to the defect, it becomes a defect operator $\hat{\bm{\phi}}$ with the same O(3) quantum number. When the defect remains conformal, $\hat{\bm{\phi}}$ exhibits power-law correlations along the defect line, with an exponent $\hat{\Delta}_{\bm{\phi}}$.  Formally, the bulk and defect operators are related by the bulk-to-defect operator-product expansion~\footnote{Here for simplicity, we suppress the $\theta$ dependence controlled by the transverse spin, as well as other less relevant terms in the expansion.}:
\begin{equation}
    \bm{\phi}(r,\theta, x_\perp)\sim r^{\hat{\Delta}_{\bm{\phi}}-\Delta_{\bm{\phi}}} \hat{\bm{\phi}}(x_\perp).
\end{equation}
 The fitted exponent  from \eqref{eq:CsLfit} provides a numerical estimate of $\hat{\Delta}_{\bm{\phi}}$. 

\begin{figure}[ht]
    \centering
    \includegraphics[width=0.9\linewidth]{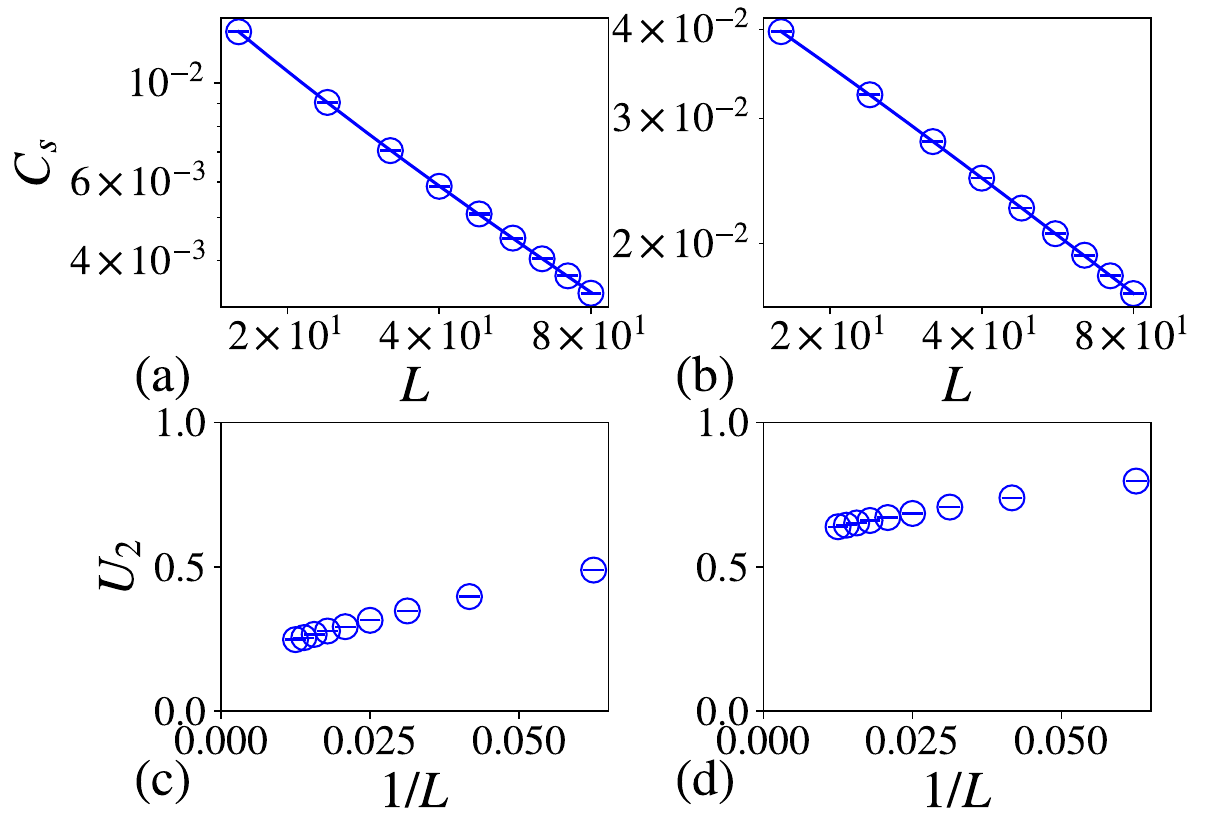}
    \caption{Finite-size scaling of observables for ordinary and special R\'enyi defects at $n=2$ in the columnar dimerized Heisenberg model. Panels (a) and (c) correspond to the ordinary defect, while panels (b) and (d) correspond to the special defect. Panels (a) and (b) show the spin correlation $C_s(L)$ on the R\'enyi line defects, while panels (c) and (d) show the Binder cumulant $U_2(L)$. System sizes range from $L=16$ to $80$. Solid lines are fits described in the text.}
    \label{fig:ord/spe}
\end{figure}

\textit{\color{blue}Ordinary and special R\'enyi defects---}
We first analyze R\'enyi defects corresponding to ordinary and special cuts. Fig.~\ref{fig:ord/spe} presents the spin correlation function and Binder cumulant for representative ordinary and special cuts at $n = 2$.

For both universality classes, the correlation function decreases with increasing system size and extrapolates to zero in the thermodynamic limit. The data are well fitted by Eq.~\eqref{eq:CsLfit}. This indicates the absence of long-range magnetic order on the R\'enyi defect.

The Binder cumulant $U_2$ decreases with system size and approaches a value smaller than one. These results confirm that ordinary and special R\'enyi defects remain disordered for all studied finite R\'enyi indices within our accessible system sizes. We do not address the behavior in the limit $n \to \infty$ in the main text. Additional data for the large-$n$ limit, obtained from $n=2L$, are presented in the SM~\cite{supplemental}.

To quantify the scaling behavior in the disordered regime, we extract the scaling dimension $\hat{\Delta}_{\bm{\phi}}$ from finite-size scaling for different R\'enyi indices. The results for $n = 2$ and $n = 3$ are summarized in Table~\ref{tab:exponent}. The table also includes results for the extraordinary cut, which will be discussed below. We have also verified that the fits are stable with respect to different choices of lower size cutoff, see SM~\cite{supplemental} for more details. The distinct values of $\hat{\Delta}_{\bm{\phi}}$ across different cuts indicate that the corresponding R\'enyi defects belong to different universality classes. 
We have further verified that ordinary and special R\'enyi defects exhibit consistent scaling behavior in the bilayer Heisenberg model, as shown in the SM~\cite{supplemental}, supporting the universality of these results.

\begin{table}[ht]
\caption{Scaling dimension $\hat{\Delta}_{\bm{\phi}}$ extracted from finite-size scaling of $C_s(L)$ for ordinary, special, and extraordinary R\'enyi defects at $n = 2$ and $n = 3$. The extraordinary defect remains in the disordered regime for these values of $n$.}
\begin{ruledtabular}
    \begin{tabular}{c c c c}
    $n$ & Ordinary & Special & Extraordinary \\
    \hline
    $2$ & $0.358(5)$ & $0.278(2)$ & $0.317(2)$ \\
    $3$ & $0.324(6)$ & $0.154(1)$ & $0.178(1)$ \\
    \end{tabular}
\end{ruledtabular}
\label{tab:exponent}
\end{table}

\begin{figure}[ht]
    \centering
    \includegraphics[width=1.0\linewidth]{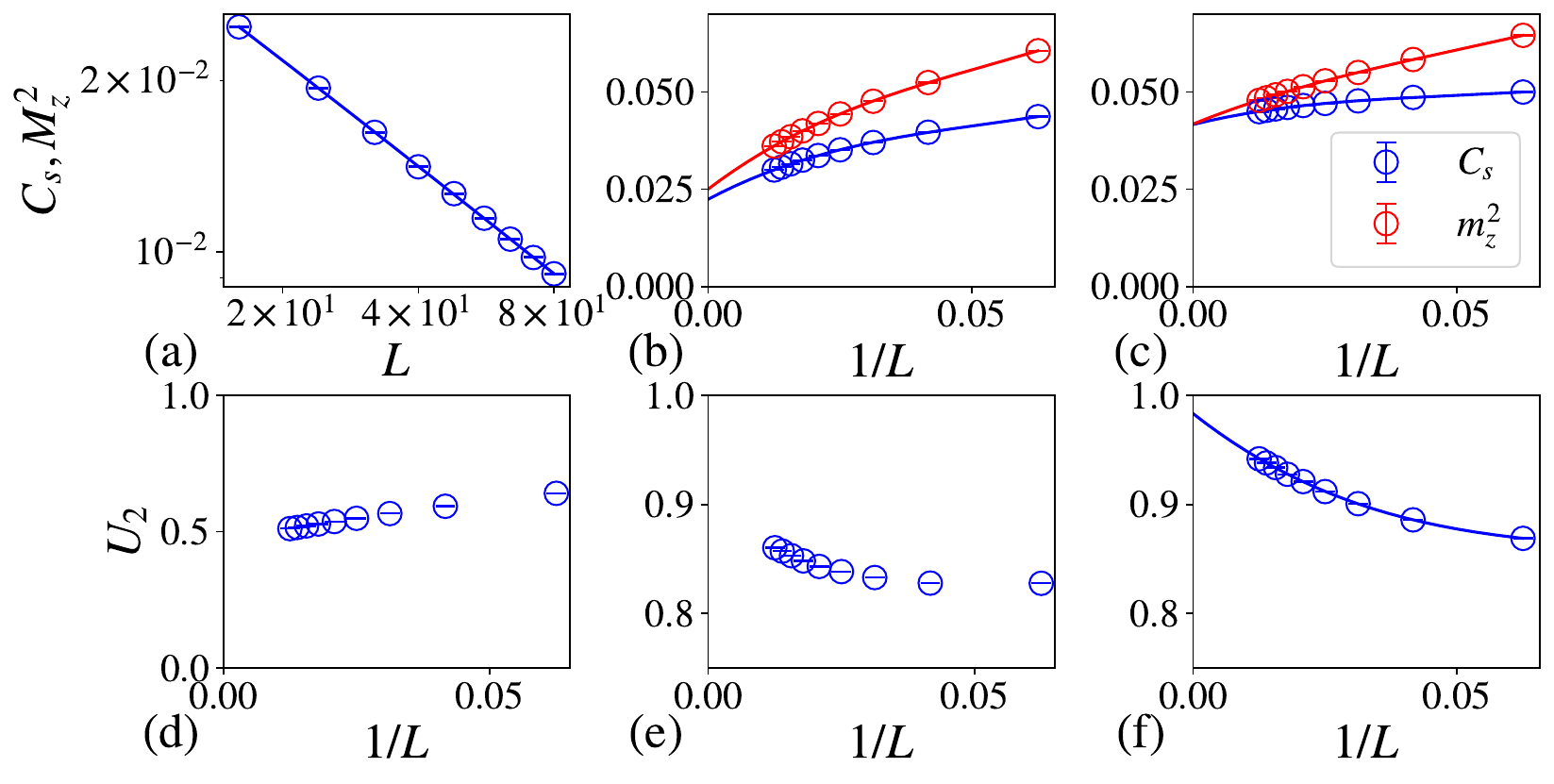}
    \caption{Finite-size scaling of observables for the extraordinary R\'enyi defect at different R\'enyi indices. Panels (a)--(c) correspond to $n=2$, $n=4$, and $n=10$, respectively, and show the spin correlation $C_s(L)$ on the R\'enyi defect line. Panels (d)--(f) show the corresponding Binder cumulant $U_2(L)$. For $n=4$ and $n=10$, panels (b) and (c) additionally show the squared magnetization $M_z^2$. System sizes range from $L=16$ to $80$. Solid lines are fits described in the text.}
    \label{fig:ext}
\end{figure}

\begin{figure}[ht]
    \centering
    \includegraphics[width=1.0\linewidth]{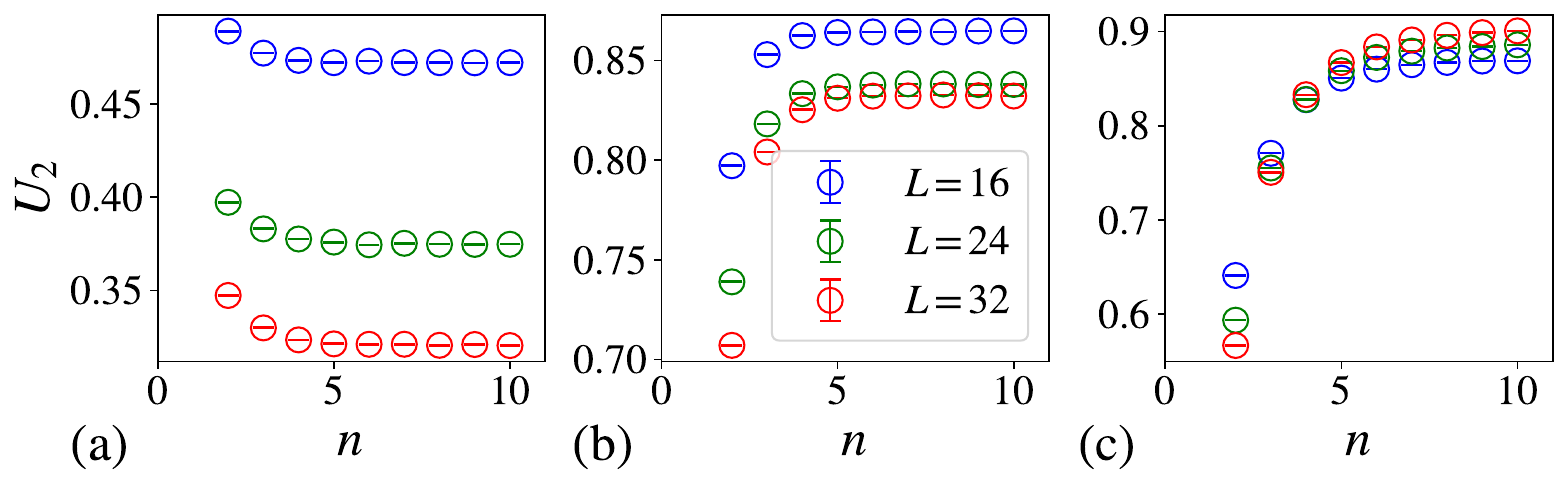}
    \caption{Binder cumulant $U_2$ as a function of R\'enyi index $n$ for (a) ordinary, (b) special, and (c) extraordinary R\'enyi defects in the columnar dimerized Heisenberg model. Different symbols correspond to system sizes $L = 16, 24, 32$. For the ordinary and special cuts, $U_2$ evolves smoothly with $n$ and no crossing is observed, consistent with the absence of a finite-$n$ phase transition. In contrast, for the extraordinary cut the Binder cumulant curves cross near $n_c \approx 3\text{--}4$, indicating a transition between a disordered regime at small $n$ and an ordered regime at larger $n$. }
    \label{fig:binder}
\end{figure}

\textit{\color{blue}Extraordinary R\'enyi defects---} 
We now turn to the extraordinary R\'enyi defect, realized in the columnar dimerized model by a tilted entanglement cut. Fig.~\ref{fig:ext} presents the spin correlation functions, squared magnetization and Binder cumulant for representative values $n = 2, 4, 10$, while Fig.~\ref{fig:binder} shows the Binder cumulant as a function of $n$ for different system sizes.

For small R\'enyi indices, such as $n = 2$, the correlation function decreases with increasing system size and extrapolates to zero in the thermodynamic limit. The data are well fitted by the form $C_s(L) = a L^{-2\hat{\Delta}_{\bm{\phi}}}(1 + b L^{-1})$, with $\hat{\Delta}_{\bm{\phi}}$ given in Table \ref{tab:exponent}. The same disordered behavior is observed at $n = 3$, where $C_s(L)$ continues to decay and shows  no indication of long-range order (see SM~\cite{supplemental}).

As the R\'enyi index increases, the Binder cumulant $U_2(L)$ grows systematically and exhibits a clear crossing near $n_c \approx 3\text{--}4$ for system sizes ranging from $L=16$ to $L=32$, as shown in Fig.~\ref{fig:binder}. Here the R\'enyi index $n$ serves as a control parameter tuning the effective state of the defect. This crossing indicates a transition between a disordered and an ordered regime at a finite critical R\'enyi index $n_c \approx 3\text{--}4$ in the thermodynamic limit. The rapid evolution of both $C_s(L)$ and $U_2(L)$ across this range of $n$ supports the presence of a true transition rather than a smooth crossover.

For intermediate R\'enyi indices, exemplified by $n = 4$, the system lies close to the transition. The decay of $C_s(L)$ is suppressed and $U_2(L)$ increases with system size, consistent with the behavior observed near the crossing point. Polynomial extrapolations of both $C_s(L)$ and $M_z^2$ yield finite intercepts, $C_s(\infty)=0.02244(10)$ and $M_z^2(\infty)=0.02506(8)$, indicating that the system is very close to, and likely on the ordered side of, the transition. The difference between the two extrapolated values suggests that finite-size corrections remain important in this near-critical regime.

For larger R\'enyi indices, exemplified by $n = 10$, the behavior changes qualitatively. The correlation function saturates to a finite value in the thermodynamic limit. A polynomial extrapolation yields consistent finite values $C_s(\infty) = 0.04165(11)$ and $M_z^2(\infty) = 0.04177(10)$ within uncertainties. The Binder cumulant approaches $U_2(\infty) = 0.9836(7)$, consistent with spontaneous symmetry breaking on the R\'enyi defect. This provides clear evidence for magnetic order on the extraordinary R\'enyi defect line.

\textit{\color{blue}Discussions---} It is useful to compare the numerical results with analytic studies of R\'enyi
defects in the interacting O$(N)$ CFT, performed in~\cite{metlitski2009entanglement} using both large-$N$ approximation and $\epsilon$ expansion. We will briefly review the results, focusing mostly on the large-$N$ solution. 

In the large-$N$ limit, the bulk is the O$(N)$ Wilson-Fisher CFT (while we have $N=3$ in the numerical simulations). 
 From solving the large-$N$ gap equation, one finds that for all $n>1$ there is a \emph{unique} physical solution with $\hat{\Delta}_{\bm{\phi}}<\Delta_{\bm{\phi}}$. In addition, $\hat{\Delta}_{\bm{\phi}}$ decreases monotonically with $n$, and approaches $0$ as $n\rightarrow \infty$. Some representative values from the large $N$ solution are  $\hat{\Delta}_{\bm{\phi}}\approx 0.34$ for $n=2$, and $0.23$ for $n=3$~\cite{metlitski2009entanglement, MClargeN}. 

Our lattice numerical results point to multiple distinct universality classes of R\'enyi-defect criticality. This suggests a richer structure of defect criticality at finite $N$ beyond the large-$N$ analysis. It  is worth pointing out that all three types of R\'enyi defects share certain qualitative features (e.g. $\hat{\Delta}_{\bm{\phi}}<1/2$ and decreases with $n$) as the large-$N$ solution. We notice that a RG calculation using $\epsilon$ expansion~\cite{metlitski2009entanglement} finds two fixed points, one stable and one unstable, when $n$ is sufficiently close to 1.

The finite-$n$ ordering transition of the extraordinary defect is
reminiscent of the behavior found near four spacetime dimensions. More precisely, for
codimension-two R\'enyi defects in general spacetime dimension $d$, the
large-$N$ analysis finds that, when $d$ is close to $4$, there is a
critical index $n_c$: below $n_c$ the defect universality class is
smoothly connected to the $D=3$ solution, while above $n_c$ the defect
develops an ordering tendency, reflected in a parametrically small
$\hat{\Delta}_{\bm{\phi}}$~\cite{metlitski2009entanglement}. This result can also be seen from the defect RG analysis.

Universal subleading contributions to the R\'enyi EE can be viewed as global observables of the underlying R\'enyi defect theory. For a bipartition containing a corner of opening angle $\theta$, the R\'enyi EE contains a universal logarithmic contribution, $S^{(n)}=aL-s_n(\theta)\log L$,
where the corner function $s_n(\theta)$ is universal data associated with the infrared R\'enyi-defect universality class and the corner geometry. Consequently, different microscopic entanglement cuts that flow to distinct ordinary, special, or extraordinary defect fixed points can have different corner functions. This provides a defect-CFT interpretation of the cut-dependent logarithmic coefficients observed in Ref.~\cite{zhao2022scaling}. In the nearly smooth limit $\theta\to\pi$, the leading behavior of $s_n(\theta)$ is controlled by the defect central charge $C_D$ (i.e. the normalization of the displacement-operator two-point function), making the connection to local defect-CFT data particularly explicit.

For smooth cylindrical bipartitions in dimerized Heisenberg models, the universal constant term in $S^{(2)}=aL+\gamma$
was likewise found to depend strongly on the microscopic cut, with cuts associated with ordinary, special, and extraordinary surface types producing different values of $\gamma$~\cite{wang2026universal}.  Establishing a quantitative relation between $\gamma$ and local defect-CFT data remains an important open problem.

The behavior in the von Neumann limit $n \to 1$ is a particularly interesting open question. Although one may expect the distinction between different R\'enyi defects to become less sharp as the replica geometry approaches a single sheet, the von Neumann EE is obtained from the derivative of the defect free energy at $n=1$. Therefore the manner in which the different defect universality classes approach the identity defect can still leave nontrivial signatures in von Neumann EE.  Since our simulations access integer $n>1$, resolving this issue would require either a controlled analytic continuation in $n$ or a different numerical approach. 
We therefore leave this question as an important direction for future work.

For the extraordinary cut, the finite-$n$ transition has a direct implication for the entanglement properties. Since the R\'enyi index can be viewed as an effective inverse temperature for the entanglement Hamiltonian, increasing $n$ gives stronger weight to the low-entanglement-energy sector without changing the entanglement spectrum itself. The transition near $n_c$ therefore indicates that this reweighting can qualitatively change the long-distance correlations probed by $\rho_A^n$: above $n_c$ long-range order develops. In this sense, different R\'enyi indices probe qualitatively different correlation structures encoded in the same $\rho_A$, providing a more direct entanglement interpretation of the transition beyond the change in defect free energy. Even below the transition, the strong $n$-dependence of $\hat{\Delta}_{\bm{\phi}}$ indicates that the extraordinary cut is not described by an $n$-independent boundary correction.

Relatedly, the ordered phase on the extraordinary R\'enyi defect found in our simulations may appear surprising, since general arguments in relativistic field theories rule out spontaneous breaking of  a continuous symmetry on line defects under reasonable assumptions about the defect RG flow~\cite{cuomo2024spontaneous}. However, in our lattice simulations, due to the sublattice structure of $M_z$, the order on the extraordinary R\'enyi defect can be interpreted as a ferromagnetic one. It is well-known that such ordering can occur in the ground states of one-dimensional quantum spin chains (for a recent discussion, see~\cite{Wang2025CombChain}). Hence it is possible that the ordering here is tied to microscopic one-dimensional degrees of freedom localized near the defect. Another possibility, suggested by the large-$N$ analysis, is that the apparent ordered phase is a finite-size manifestation of the defect order parameter with a very small $\hat{\Delta}_{\bm{\phi}}$.

\textit{\color{blue}Conclusion---} In this Letter, we have numerically studied R\'enyi defects at the (2+1)$D$ O(3) QCP by measuring local observables on the replica spacetime manifold. Microscopically different entanglement bipartitions realize ordinary, special, and extraordinary R\'enyi defects, which exhibit distinct critical behavior as exemplified by the scaling dimensions. Most notably, the extraordinary R\'enyi defect shows a finite-$n$ transition from a disordered phase at small R\'enyi index $n$ to a ferromagnetically ordered phase at larger $n$, while the ordinary and special defects remain disordered within the accessible range of $n$. These results show that the sensitivity of entanglement observables to microscopic cut geometry can be understood in terms of distinct defect universality classes in the IR.

\begin{acknowledgments}
\textit{\color{blue}
Acknowledgments}\,---\,  We thank Yanliang Guo, Zenan Liu and Yan-Cheng Wang for helpful discussions. MC acknowledges enlightening conversations with Yichul Choi, Max Metlitski, Yifan Wang and Xiao-Chuan Wu, and collaboration with Nayan Myerson-Jain and G. Shankar on related topics. MC is also grateful for the hospitality from the Institute for Advanced Study. This project is supported by the Scientific Research Project (No. WU2025B011), the Feng-Ying Career Development Chair Fund and the Start-up Funding of Westlake University, and Quantum MetaKey (QuaMeK) computing. The authors thank the IT service office and the high-performance computing center of Westlake University.

Y. Z. and Z. W. contributed equally to this work.
\end{acknowledgments}

\textit{\color{blue}Data availability---} The data supporting the findings of this article are openly available at \cite{data}.

\bibliography{main}

\clearpage
\onecolumngrid

\setcounter{page}{1}
\renewcommand{\thepage}{S\arabic{page}}

\setcounter{section}{0}
\setcounter{equation}{0}
\setcounter{figure}{0}
\setcounter{table}{0}

\renewcommand{\thesection}{S\arabic{section}}
\renewcommand{\theequation}{S\arabic{equation}}
\renewcommand{\thefigure}{S\arabic{figure}}
\renewcommand{\thetable}{S\arabic{table}}

% ====== SM Title block ======

\begin{center}
{\large\bfseries Criticality on R\'enyi defects at (2+1)$d$ O(3) quantum critical points}\\[0.5em]
{\large\bfseries -Supplemental Material-}
\vspace{1em}

Yanzhang Zhu,$^{1,2,3}$ Zhe Wang,$^{1,2}$ Meng Cheng,$^{4}$ and Zheng Yan$^{1,2}$\\[0.5em]
{\small
$^{1}$Department of Physics, School of Science and Research Center for
Industries of the Future, Westlake University, Hangzhou 310030, China\\
$^{2}$Institute of Natural Sciences, Westlake Institute for Advanced Study,
Hangzhou 310024, China\\
$^{3}$State Key Laboratory of Surface Physics and Department of Physics,
Fudan University, Shanghai 200433, China\\
$^{4}$Department of Physics, Yale University, New Haven, Connecticut 06511, USA
}\\[0.5em]
\end{center}

\vspace{1.0em}
% ============================
\section{Universality of ordinary and special R\'enyi defects}

We consider the spin-$1/2$ bilayer Heisenberg model,
\begin{equation}
H = J \sum_{\langle ij \rangle, \ell} \mathbf{S}_{i,\ell} \cdot \mathbf{S}_{j,\ell}
+ J' \sum_{i} \mathbf{S}_{i,1} \cdot \mathbf{S}_{i,2},
\end{equation}
where $\ell = 1,2$ labels the two layers, $J$ is the intralayer coupling, and $J'$ is the interlayer coupling. The system is tuned to its bulk quantum critical point at $J'/J = 2.522$.

We construct R\'enyi defects using the same replica approach as in the main text by performing an entanglement bipartition of a system with periodic boundary conditions. The resulting R\'enyi defect line is defined along the entanglement cut.

\begin{figure}[ht]
    \centering
    \includegraphics[width=0.5\linewidth]{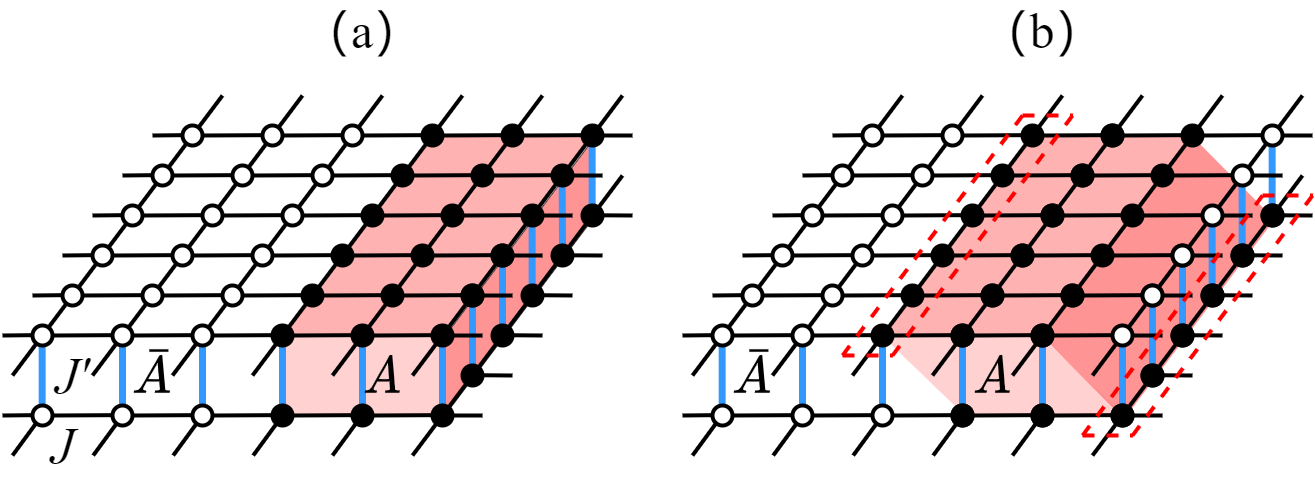}
    \caption{Schematic illustration of the bilayer Heisenberg model and entanglement bipartitions realizing (a) ordinary and (b) special R\'enyi defects. The system has periodic boundary conditions and is divided into the subsystem $A$ (red region) and the environment $\bar A$ by the entanglement cut, shown here by black and white spins. Thin black bonds denote intralayer couplings $J$, and thick blue bonds denote interlayer couplings $J'$. In (a), the bipartition does not break interlayer singlets, and observables are measured along the boundary of subsystem $A$, corresponding to the ordinary defect. In (b), the bipartition effectively cuts through the interlayer dimers along a tilted direction, leaving a chain of unpaired spins along the defect; the measurement region is indicated by red dashed boxes, corresponding to the special defect.}
    \label{fig:model_bilayer}
\end{figure}

Different choices of bipartition in the bilayer geometry realize ordinary and special R\'enyi defects, as illustrated in Fig.~\ref{fig:model_bilayer}. The system is divided into two subsystems by the entanglement cut, and observables are measured along the boundary of subsystem $A$. 

For the ordinary cut, the bipartition does not break the interlayer singlets, and no additional low-energy degrees of freedom are generated along the defect. In contrast, for the special cut, the bipartition effectively cuts through the interlayer dimers along a tilted direction, leaving a chain of unpaired spins along the defect. The measurement region in this case is indicated in Fig.~\ref{fig:model_bilayer}(b). This construction is directly analogous to the special surface transition in the dimerized Heisenberg model, where cutting strong bonds produces dangling spin chains. The highlighted chain in Fig.~\ref{fig:model_bilayer}(b) indicates the defect spins used to measure $C_s(L)$ and $U_2(L)$.

\begin{figure}[ht]
    \centering
    \includegraphics[width=0.5\linewidth]{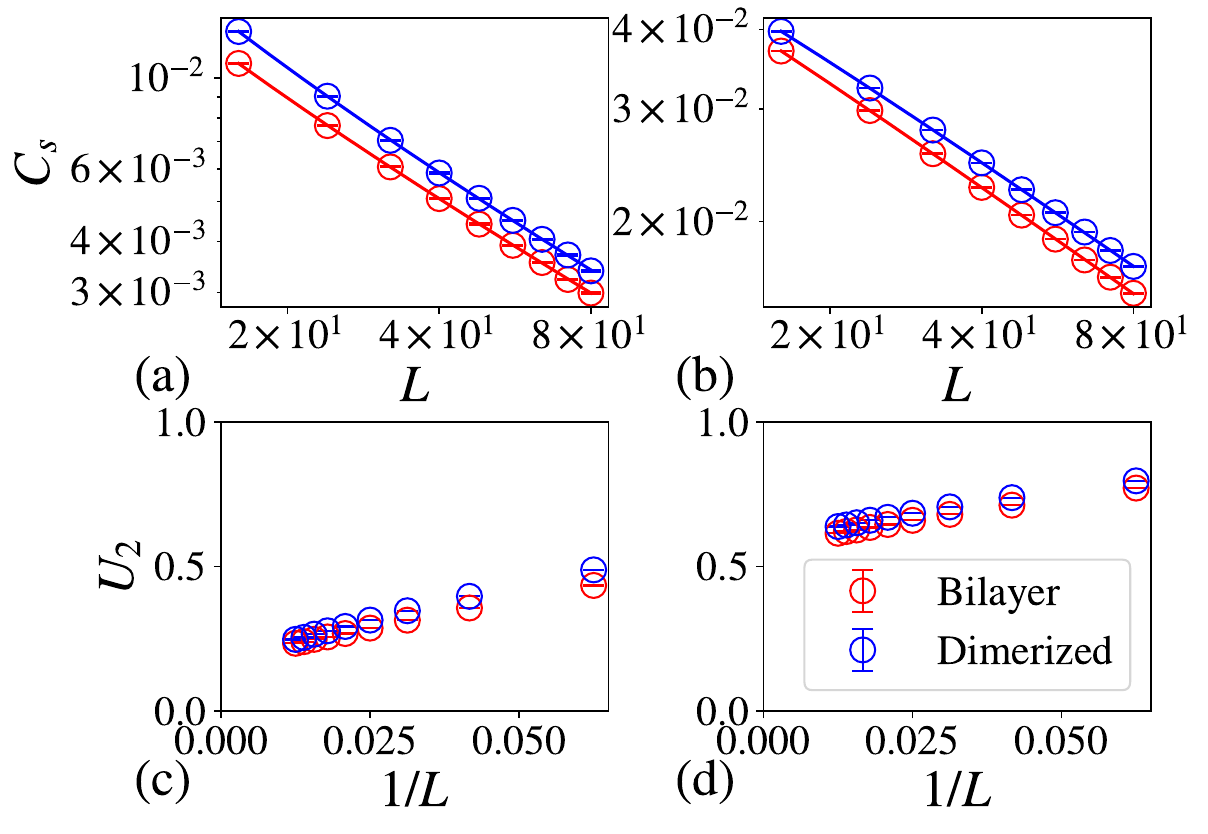}
    \caption{Comparison of finite-size scaling between the bilayer (red) and columnar dimerized (blue) Heisenberg models at R\'enyi index $n=2$. Panels (a)(b) show the surface correlation $C_s(L)$ for the ordinary and special defects, respectively, while panels (c)(d) show the corresponding Binder cumulant $U_2(L)$. In both cases, the bilayer and dimerized data exhibit very similar scaling behavior: $C_s(L)$ decreases with system size and $U_2(L)$ approaches a value smaller than one.}
    \label{fig:bilayer}
\end{figure}

We compute the spin correlation $C_s(L)$ and Binder cumulant $U_2(L)$ at R\'enyi index $n=2$ for system sizes up to $L=80$. Fig.~\ref{fig:bilayer} compares the finite-size scaling of these observables between the bilayer and columnar dimerized models for both ordinary and special cuts.

Fitting the correlation function using the same scaling form as in the main text yields scaling dimensions $\Delta_{\rm ord} = 0.353(5)$ and $\Delta_{\rm spec} = 0.281(2)$ for the bilayer model, which are consistent within uncertainties with the corresponding values in the columnar dimerized model.

These results support the universality of ordinary and special R\'enyi defects, indicating that their scaling behavior is insensitive to microscopic details of the model.

\section{Universality of extraordinary R\'enyi defects}

We consider the spin-$1/2$ staggered dimerized Heisenberg model,
\begin{equation}
H = J \sum_{\langle ij\rangle} \mathbf{S}_i \cdot \mathbf{S}_j 
+ J' \sum_{\langle ij\rangle'} \mathbf{S}_i \cdot \mathbf{S}_j ,
\end{equation}
where $\langle ij\rangle$ and $\langle ij\rangle'$ denote weak and strong bonds, respectively. The system is tuned to its bulk quantum critical point at $J'/J = 2.5196$.

\begin{figure}[ht]
    \centering
    \includegraphics[width=0.15\linewidth]{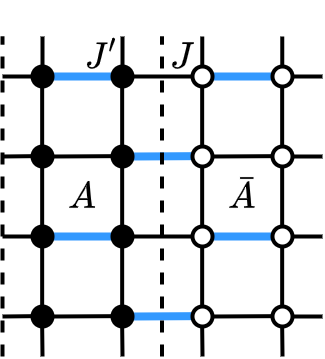}
    \caption{Schematic illustration of the staggered dimerized Heisenberg model and the entanglement bipartition realizing the extraordinary R\'enyi defect. Thin black bonds denote weak couplings $J$, and thick blue bonds denote strong couplings $J'$. The system is divided into subsystems $A$ (black) and its environment $\bar{A}$ (white) by the entanglement cut.
}
    \label{fig:model_staggered}
\end{figure}

We construct the R\'enyi defect using the same entanglement bipartition and replica approach as in the main text. An illustration of the entanglement cut is shown in Fig.~\ref{fig:model_staggered}.

Along the entanglement cut, the bipartition intersects bonds in an alternating sequence of weak and strong couplings. This structure is distinct from the ordinary case, where dimers remain intact, and the special case, where strong bonds are cut to produce dangling spins. The alternating pattern provides a microscopic realization of the extraordinary R\'enyi defect.

\begin{figure}[ht]
    \centering
    \includegraphics[width=0.5\linewidth]{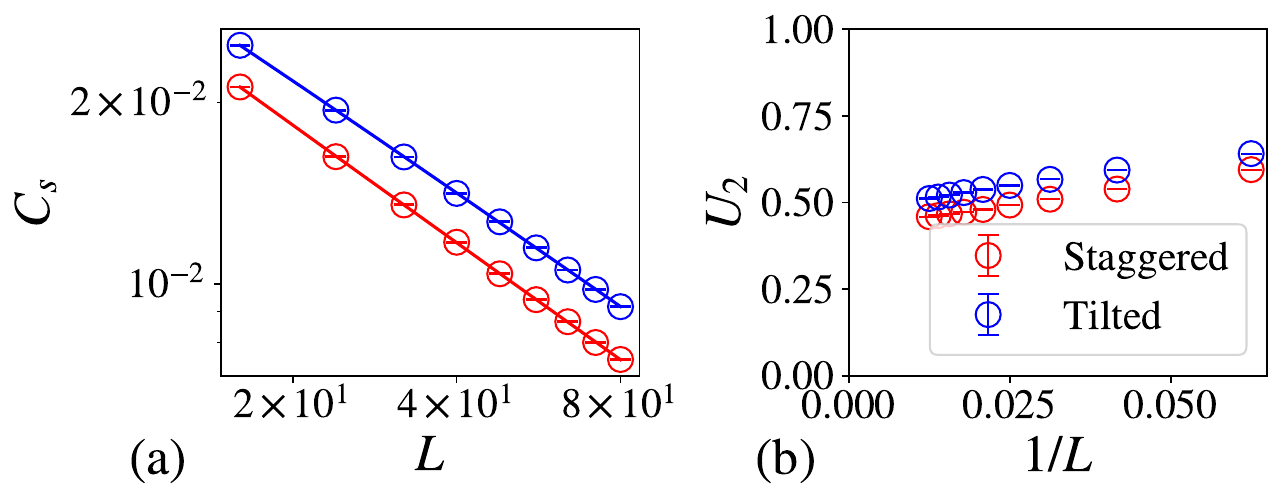}
    \caption{Comparison of finite-size scaling between the staggered (red) and tilted columnar (blue) dimerized Heisenberg models at R\'enyi index $n=2$ for the extraordinary defect. Panel (a) shows the surface correlation $C_s(L)$, and panel (b) shows the Binder cumulant $U_2(L)$. The two models exhibit very similar scaling behavior: $C_s(L)$ decreases with system size and $U_2(L)$ approaches a value smaller than one.
}
    \label{fig:staggered}
\end{figure}

We compute the surface correlation $C_s(L)$ and Binder cumulant $U_2(L)$ at R\'enyi index $n=2$ for system sizes up to $L=80$. Fig.~\ref{fig:staggered} shows the corresponding results for the staggered model, together with those obtained from the tilted columnar construction.

Fitting the correlation function using the same scaling form as in the main text yields scaling dimensions $\Delta_{\rm stag} = 0.321(3)$ for the staggered model that agree within uncertainties with those obtained from the tilted columnar construction. These results demonstrate that the extraordinary R\'enyi defect is characterized by the same critical exponent in both realizations, despite differences in microscopic geometry.

\section{R\'enyi-$n=3$ data}
\begin{figure}[ht]
    \centering
    \includegraphics[width=0.7\linewidth]{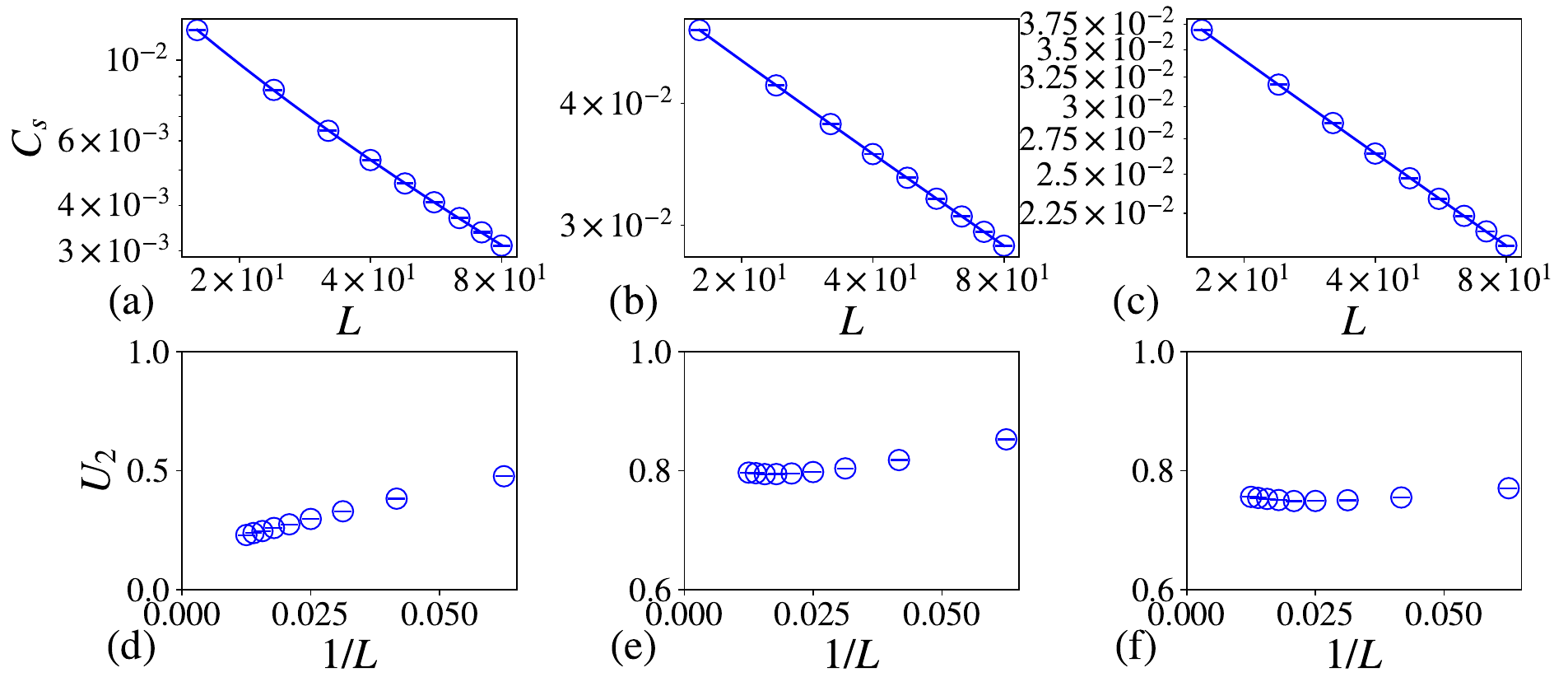}
    \caption{Finite-size scaling of the R\'enyi-$n=3$ defect for the three types of cuts discussed in the main text. Panels (a)–(c) show the surface correlation $C_s(L)$ for the ordinary, special, and extraordinary defects, respectively, while panels (d)–(f) show the corresponding Binder cumulant $U_2(L)$. The data provide the basis for the extraction of scaling dimensions reported in the main text.}
    \label{fig:renyi3}
\end{figure}

We present the finite-size scaling of the R\'enyi-$n=3$ defect for the ordinary, special, and extraordinary cases in Fig.~\ref{fig:renyi3}. The surface correlation $C_s(L)$ and Binder cumulant $U_2(L)$ are computed using the same setup as in the main text. The data shown here are used to extract the scaling dimensions reported in the main text.

\section{Stability of scaling fits}

To assess the robustness of the extracted scaling dimensions, we perform fits using different lower size cutoffs $L_{\min}$ while keeping the largest system size fixed at $L_{\max}=80$. The results for the three types of R\'enyi defects at $n=2$ are summarized in Table~\ref{tab:stability}.

\begin{table}[ht]
\caption{Stability of the scaling dimension $\Delta$ extracted from fits of $C_s(L)$ at R\'enyi index $n=2$ using different lower size cutoffs $L_{\min}$, with $L_{\max}=80$ fixed. The last row gives the reduced chi-squared values for the fits with $L_{\min}=16$.}
    \begin{tabular*}{0.6\columnwidth}{@{\extracolsep{\fill}}c c c c}
    \hline 
    \hline
    $L_{min}$ & Ordinary & Special & Extraordinary \\
    \hline
    $16$ & $0.358(5)$ & $0.278(2)$ & $0.317(2)$ \\
    $24$ & $0.360(9)$ & $0.276(3)$ & $0.319(4)$ \\
    $32$ & $0.38(2)$ & $0.280(5)$ & $0.328(7)$ \\
    $40$ & $0.40(3)$ & $0.287(9)$ & $0.34(2)$ \\
    \hline
    $\chi^2_{\rm red}$ & $1.035$ & $1.178$ & $1.606$ \\
    \hline 
    \hline
    \end{tabular*}
\label{tab:stability}
\end{table}

The fitted values remain stable within uncertainties for all three cases. A mild systematic drift is observed for the ordinary defect as $L_{\min}$ increases, which we attribute to residual finite-size effects. The $\chi^2_{\rm red}$ values for the $L_{\min}=16$ fits are close to unity as well, supporting the quality of the scaling fits. Overall, the results support the reliability of the scaling dimensions reported in the main text.

\section{Behavior of R\'enyi defects in the large-$n$ limit}
\begin{figure}[ht]
\centering
\includegraphics[width=0.7\linewidth]{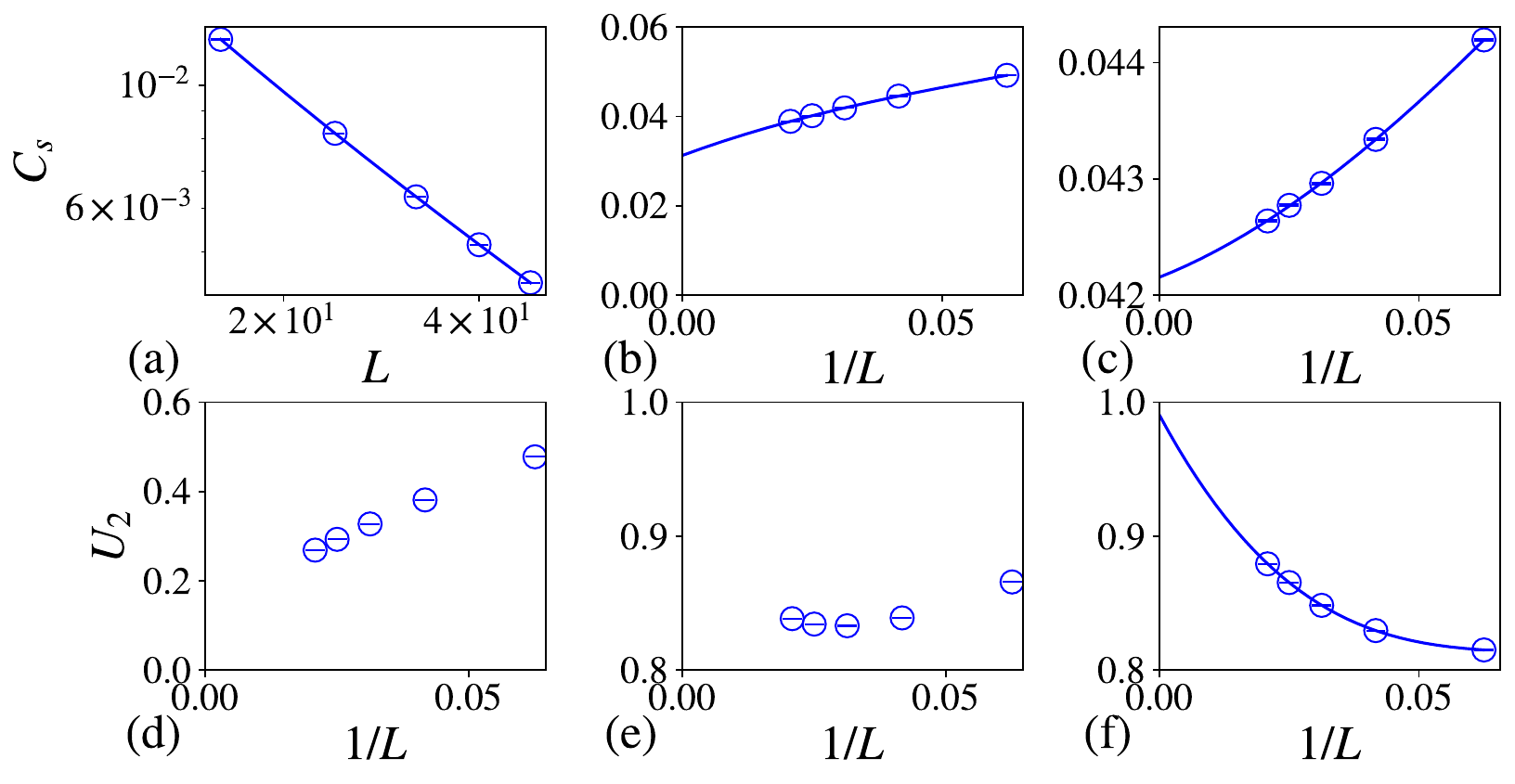}
\caption{Finite-size scaling of observables for R\'enyi defects in the large-$n$ limit, realized by $n=2L$. Columns (left to right) correspond to the ordinary, special, and extraordinary defects. Top row shows the spin correlation $C_s(L)$ on the R\'enyi defect line, and bottom row shows the Binder cumulant $U_2(L)$. For the ordinary defect, $C_s(L)$ decreases with increasing system size and extrapolates to zero, while $U_2(L)$ remains small, indicating a disordered regime. For the extraordinary defect, $C_s(L)$ extrapolates to a finite nonzero value and $U_2(L)$ approaches unity as $L \to \infty$, indicating spontaneous symmetry breaking on the R\'enyi defect. For the special defect, $C_s(L)$ shows a tendency toward a finite value, while $U_2(L)$ approaches a value smaller than one, indicating that the behavior in this regime remains subtle within the accessible system sizes. System sizes range from $L = 16$ to $48$. Solid lines are fits described in the text.}
\label{fig:renyi2l}
\end{figure}

To further probe the behavior of R\'enyi defects beyond the finite-$n$ regime discussed in the main text, we examine the large-$n$ limit by considering $n=2L$, which corresponds to an effective low-temperature limit of the entanglement Hamiltonian. The results are shown in Fig.~\ref{fig:renyi2l}.

For the ordinary defect, the correlation function $C_s(L)$ decreases with system size and extrapolates to zero, while the Binder cumulant remains small, consistent with a disordered regime. In contrast, for the extraordinary defect, polynomial extrapolation yields a finite value $C_s(\infty)=0.0421(2)$ and $U_2(\infty)=0.990(2)$, providing clear evidence of long-range order on the defect.

The behavior of the special defect is more subtle. While $C_s(L)$ shows a tendency toward a finite value, with polynomial extrapolation giving $C_s(\infty)=0.0312(1)$, the Binder cumulant approaches a non-unit value and shows no clear convergence to an ordered limit within the accessible system sizes. Combined with the finite-$n$ results in the main text, where the special defect remains disordered for $n=2$ to $10$, we conclude that there is no clear evidence for a finite-$n$ transition in the special case within the accessible range.

\end{document}